\documentclass[letter,traditabstract]{aa} 
\usepackage{graphicx}
\usepackage{txfonts}
\usepackage{epsfig}
\usepackage{longtable}
\usepackage{lscape}
\usepackage{amssymb} 
\usepackage{natbib} 
\usepackage[colorlinks=true,citecolor=blue]{hyperref} 
\usepackage{refcount}

\def \igr {\mbox{IGR~J17544$-$2619}}

\def \nustar {{\em NuSTAR}}
\def \sw {{\em Swift}}

\def \hcm {\hbox {\ifmmode $ atom cm$^{-2}\else atom cm$^{-2}$\fi}}

\def \arcsec {\hbox{$^{\prime\prime}$}}

\def \ATel {ATel} 
\def \apj {ApJ}
\def \apjl {ApJL}
\def \aap {A\&A}

\def \asr {AdSpR}

\def \bcao {BCAO}

\def \mnras {MNRAS}

\def \ssr {SSRv}
\def \spie {SPIE Conf. Ser.}

\begin{document} 

\title{Giant outburst from the supergiant fast X-ray transient IGR~J17544$-$2619: accretion from a transient disc?\thanks{Tables 
\ref{sfxt14:tab:alldata}--\ref{sfxt14:tab:broadfits}, and Fig.~\ref{sfxt14:fig:broadspec} are available 
in electronic form at the CDS via anonymous ftp to {\tt cdsarc.u-strasbg.fr} (130.79.128.5) or via 
 \href{http://cdsweb.u-strasbg.fr/cgi-bin/qcat?J/A+A/vol/page}{\tt http://cdsweb.u-strasbg.fr/cgi-bin/qcat?J/A+A/vol/page}   
 } 
             } 
\titlerunning{Transient disc accretion in SFXTs}
\authorrunning{Romano et al.}
\author{P.\ Romano\inst{1}
        \and 
         E.\ Bozzo\inst{2} 
         \and
          V. Mangano\inst{3}
         \and
         P. Esposito\inst{4,5}  
          \and
         G.\ Israel\inst{6}  
         \and 
          A.\ Tiengo\inst{7,4,8} 
        \and \\
          S. Campana\inst{9}
          \and
          L.\ Ducci\inst{10,2} 
         \and 
        C. Ferrigno\inst{2} 
         \and 
         J.A.\ Kennea\inst{3}
           }
   \institute{INAF, Istituto di Astrofisica Spaziale e Fisica Cosmica - Palermo,
              Via U.\ La Malfa 153, I-90146 Palermo, Italy \\
              \email{romano@ifc.inaf.it}
               \and 
         ISDC Data Center for Astrophysics, Universit\'e de Gen\`eve, 16 chemin d'\'Ecogia, 1290 Versoix, Switzerland
            \and
         Department of Astronomy and Astrophysics, Pennsylvania State 
              University, University Park, PA 16802, USA
                \and 
         INAF, Istituto di Astrofisica Spaziale e Fisica Cosmica - Milano, 
              Via E.\ Bassini 15,  I-20133 Milano,  Italy 
                 \and 
          Harvard--Smithsonian Center for Astrophysics, 60 Garden Street,
                Cambridge, MA 02138, USA
                 \and 
         INAF, Osservatorio Astronomico di Roma, 
              Via Frascati 33, I-00040, Monte Porzio Catone, Italy 
                \and 
        Istituto Universitario di Studi Superiori, Piazza della Vittoria 15, I-27100 Pavia, Italy
                \and 
        INFN, Sezione di Pavia, via A. Bassi 6, I-27100 Pavia, Italy 
                \and
         INAF, Osservatorio Astronomico di Brera, 
              Via E.\ Bianchi 46, I-23807 Merate, Italy 
             \and 
             Institut f\"ur Astronomie und Astrophysik, Eberhard Karls Universit\"at, 
            Sand 1, 72076 T\"ubingen, Germany
             }
\date{Received 26 January 2015; accepted 12 February 2015}

\abstract{Supergiant fast X-ray transients (SFXTs) are high mass X-ray binaries 
associated with OB supergiant companions and characterised by an X-ray 
flaring behaviour whose dynamical range reaches 5 orders of magnitude on timescales of a 
few hundred to thousands of seconds. 
Current investigations concentrate on finding possible mechanisms to inhibit accretion in SFXTs
and explain their unusually low average X-ray luminosity. 
We present the {\it Swift} observations of an  
exceptionally bright outburst displayed by the SFXT IGR~J17544$-$2619 on 2014 October 10 
when the source achieved a peak luminosity of $3\times10^{38}$\,erg\,s$^{-1}$. 
This extends the total source dynamic range to $\gtrsim$10$^6$, the largest (by a factor of 10) 
recorded so far from an SFXT. 
Tentative evidence for pulsations at a period of $11.6$\,s is also reported.  
We show that these observations challenge, for the first time, the maximum 
theoretical luminosity achievable by an SFXT and propose that this giant outburst was due to  
the formation of a transient accretion disc around the compact object.}


   \maketitle

	\section{Introduction\label{sfxt14:intro}}
 
Supergiant fast X-ray transients (SFXTs) are high mass X-ray binaries (HMXBs) 
hosting a neutron star (NS) and 
an OB supergiant companion \citep{Sguera2005,Negueruela2006} 
which display X-ray flares reaching, for a few hours, 
10$^{36}$--10$^{37}$~erg~s$^{-1}$ \citep[see][for a recent review]{Romano2014:sfxts_catI}. This is 
at odds with normal supergiant HMXBs displaying a fairly constant average luminosity with typical 
variations by a factor of 10--50 on time scales of few hundred to thousands of seconds. 
SFXTs are also significantly sub-luminous with respect to classical  Sg-HMXBs like Vela X-1 \citep[][]{Bozzo2015:underluminous}, 
and show a dynamical range  up to 5 orders of magnitude, as their luminosities can be as low as $\sim 10^{32}$~erg~s$^{-1}$
during quiescence \citep[e.g.][]{zand2005,Bozzo2010:quiesc1739n08408}. 

The transient \object{IGR~J17544$-$2619} \citep{Sunyaev2003} is one of the SFXT prototypes, 
in which a NS is orbiting every 4.926$\pm$0.001\,d \citep{Clark2009:17544-2619period} an 
O9Ib star located at 3.6\,kpc \citep{Pellizza2006, Rahoui2008}. 
It  has been showing large flux swings since its discovery 
\citep{zand2004:17544bepposax, zand2005, Rampy2009:suzaku17544,Romano2014:sfxts_catI}, 
with the brightest flux recorded by \sw/XRT \citep[][]{Burrows2005:XRT} 
during the 2013 June 28 outburst \citep[][]{Romano2013:atel5179}   
at $\sim6.5\times10^{-9}$ erg\,cm$^{-2}$\,s$^{-1}$ (0.3--10\,keV, unabsorbed) 
corresponding to  $L\sim 10^{37}$\,erg\,s$^{-1}$. 
These properties 
were initially interpreted in terms of accretion onto the NS  
from a ``clumpy wind'' 
\citep[][]{zand2005}. 
However, as discussed in \citet{Bozzo2013:COSPAR_sfxt}, we now know that 
the X-ray variability of SFXTs cannot be easily reconciled with a simple extreme clumpy wind model. 
Alternatives have been proposed to explain the large X-ray  luminosity swings, invoking either the presence of 
magnetic/centrifugal barriers \citep{Grebenev2007,Bozzo2008} 
or a subsonic settling accretion regime \citep{Shakura2014:bright_flares}. 
A recent \nustar\ observation \citep[][]{Bhalerao2015:line17544}, revealed in \igr\ a  
cyclotron line at 17\,keV, yielding the first measurement of the magnetic field in an SFXT at 
$\sim 1.5\times10^{12} $\,G, as typical of accreting NS in HMXBs. 

In this Letter, we present the extraordinary set of observations collected by \sw\ 
during the 2014 October 10 outburst of \igr\ (Sect.~\ref{sfxt14:sample}), 
during which the source reached an unabsorbed mean flux of 
$\sim4.8\times10^{-8}$ erg cm$^{-2}$ s$^{-1}$ (0.3--10\,keV; Sect.~\ref{sfxt14:superburst})
thus exceeding all preceding records by a factor of 10. 
During this event evidence for a transient pulsation at $\sim 12$\,s (Sect.~\ref{sfxt14:pulsations})
and an expanding X-ray halo around the source 
were found in the XRT data.

\setcounter{figure}{0}  
\begin{figure}
\vspace{-1.1truecm}
\hspace{-0.5truecm}
 \centerline{\includegraphics[width=9.5cm,height=8.5cm,angle=0]{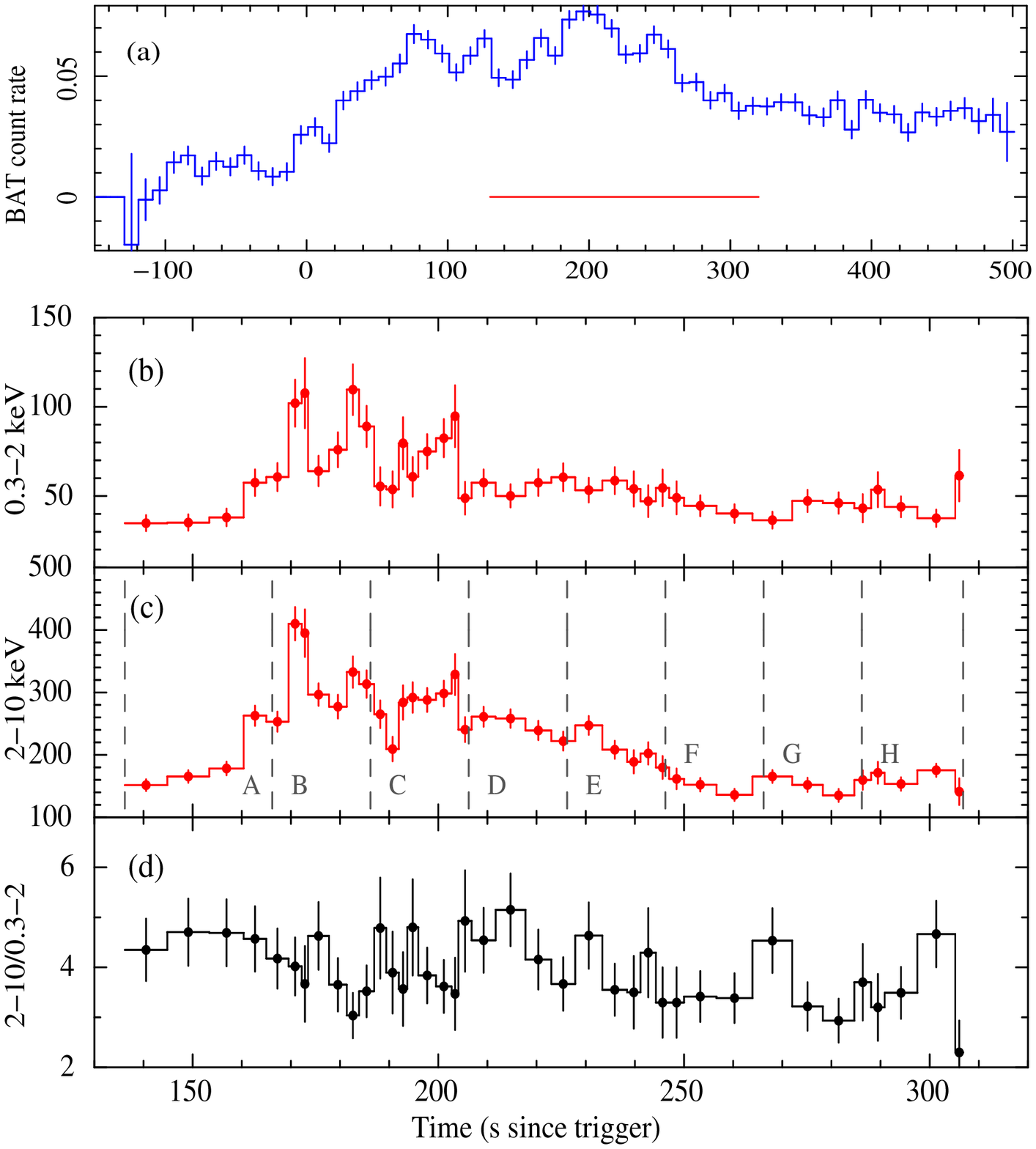}}
\vspace{-1.6truecm}
 \caption[XRT light curve]{Light curves of the 2014-10-10 outburst of IGR J17544 $-$2619. 
(a) BAT (14--150\,keV, 10\,s binning): the red horizontal line marks the XRT data span; 
(b) XRT soft band (0.3--2\,keV, at 30 counts bin$^{-1}$),  
(c) XRT hard band (2--10\,keV, at 30 counts bin$^{-1}$),  
 (d) XRT hardness ratio (2--10 keV/0.3--2 keV).   
The dashed vertical lines 
mark the time selections for simultaneous 
BAT and XRT spectroscopy (Sect.~\ref{sfxt14:superburst}). 
 		\label{sfxt14:fig:lcvs} 
}
\end{figure}

	\section{Observations and data reduction} \label{sfxt14:sample}

\igr\  triggered \sw/BAT \citep[][]{Barthelmy2005:BAT} 
on 2014 October 10 at  $T_0=$15:04:19 UT 
\citep[image trigger=614903, ][]{Romano2014:atel6566}.  
\sw\ immediately slewed to the target, 
so that the narrow-field instruments started observing at $T_0+$132\,s. 
The automated target (AT, sequence  00614903000) ran for four orbits, until $T_0+17.5$\,ks).  
Follow-up target of opportunity observations 
were obtained  (sequences 00035056161--166). 
The data cover the first 5\,d after the beginning of the outburst, 
with  6 \sw\ observations for a net XRT exposure of $\sim 12.5$\,ks (see 
Table~\ref{sfxt14:tab:alldata}). 
Based on the most recent outburst ephemeris for this source 
\citep[$P=4.92693\pm0.00036$\,d, periastron at MJD $53732.65\pm0.23$;][]{Smith2014:atel6227}, 
the outburst started at phase $\phi=0.11$.

The data were processed and analysed using the standard software ({\sc FTOOLS} v6.16),  
calibration (CALDB 20140709), and methods. 
Background-subtracted BAT light curves were created in the standard energy bands 
and mask-weighted spectra were extracted during the first orbit of data. 
The XRT data were processed and filtered with {\sc xrtpipeline} (v0.13.1). 
Pileup was corrected for (ObsID 00614903000 only) by adopting standard procedures
\citep[][]{Romano2006:060124,vaughan2006:050315}:  
source events were extracted from 
annuli  (inner/outer radii of 4/20 pix for WT, 5/30 pix for PC; 1 pixel $\sim2.36$\arcsec).  
Light curves were 
corrected for point spread function  losses, vignetting and 
background subtracted. 
For our spectral analysis, we extracted events in the same regions as 
those adopted for the light curve creation and only considered grade 0 events for WT data. 
For our timing analysis, performed on the grade 0--2 WT events, 
we also converted the event arrival times to the 
Solar system barycentre with {\sc barycorr}. 
Archival XRT data were also considered and similarly processed, from the following bright outbursts: 
2011-03-24 \citep[][$\phi=0.96$]{Farinelli2012:sfxts_paperVIII}, 
2013-06-28 \citep[][$\phi=0.86$]{Romano2013:atel5179}; 
2013-09-11 \citep[][$\phi=0.15$]{Romano2013:atel5388};  
2014-05-25 \citep[][$\phi=0.16$]{Romano2014:atel6173}.

\setcounter{figure}{2}  
\begin{figure}
\vspace{-0.3truecm}
\hspace{-0.6truecm}
\centerline{\includegraphics[width=8.8cm,height=7cm,angle=0]{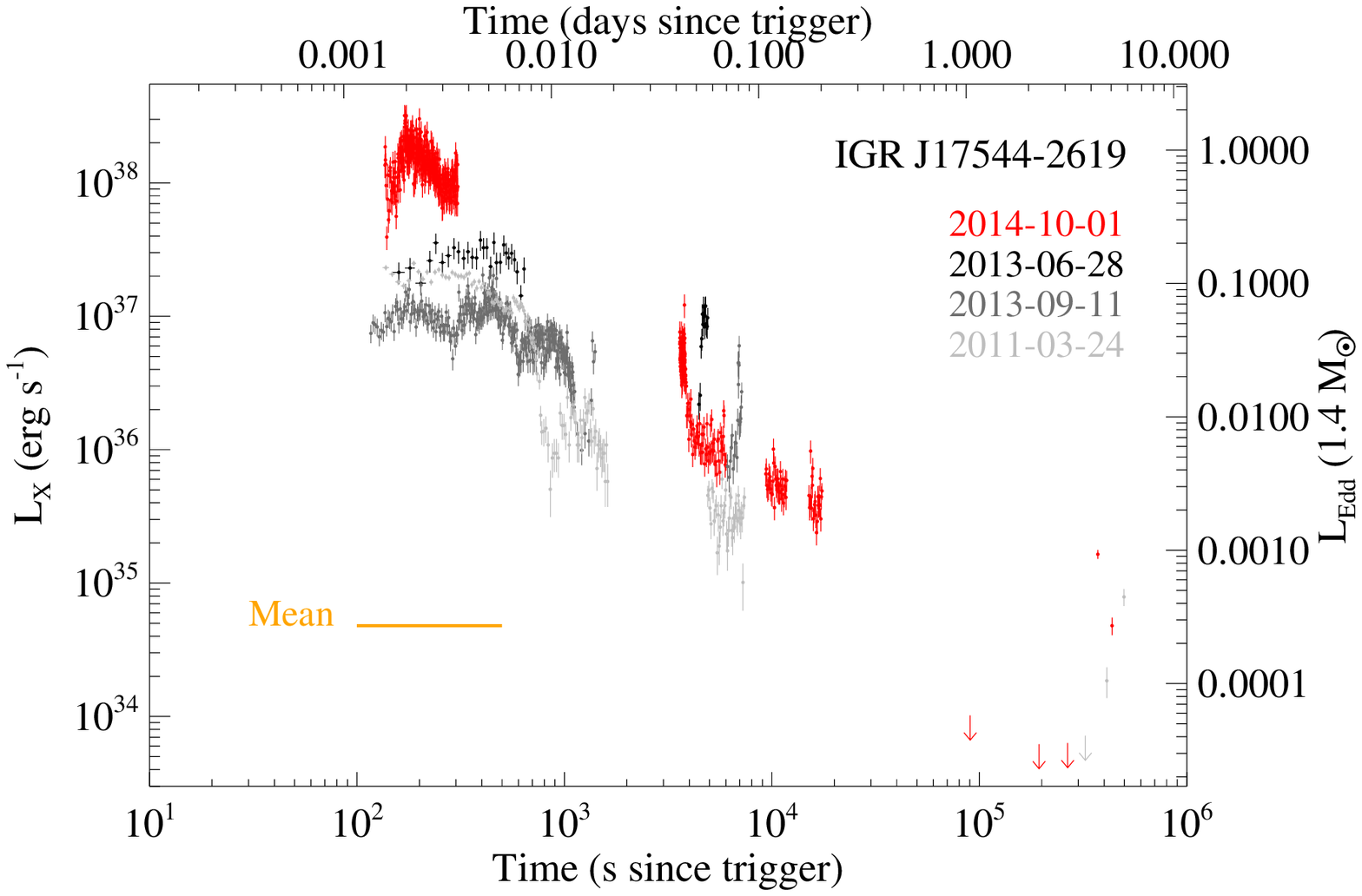}}
\vspace{-0.3truecm}
\caption[Bolometric luminosity X-ray light curve]{Bolometric X-ray luminosity light curves of 
the brightest outbursts recorded by \sw\ for this source. Red: 2014-10-10,  
light grey: 2011-03-24, 
black: 2013-06-28, 
dark grey: 2013-09-11. 
The horizontal (orange) line marks the average level for this source. 
 		\label{sfxt14:fig:lcv_lum} }
\end{figure} 

\section{Analysis and results} \label{sfxt14:results} 

\subsection{A giant outburst?} \label{sfxt14:superburst} 

 Figure~\ref{sfxt14:fig:lcvs} shows the \sw\ BAT and XRT light curves of the first orbit of data 
(before the first Earth occultation). 
While no spectral variations could be detected in the BAT data, the XRT data show 
variations in the hardness ratio. 
To investigate their nature, we extracted simultaneous XRT and BAT 
spectra in 8 time intervals (marked with A, .. , H in Fig.~\ref{sfxt14:fig:lcvs}c), 
as well as in the total  
(Orb 1, see Fig.~\ref{sfxt14:fig:broadspec}),      
and fit them in the 0.3--10\,keV and 14--70\,keV bands, respectively (see Table~\ref{sfxt14:tab:broadfits}), 
with an absorbed power law with a high energy cut-off ({\sc phabs*power*highecut} in {\sc XSPEC}). 
Even this simple spectral model does not allow us to firmly establish if variations of the 
absorption or intrinsic emission have a leading role in the hardness ratio modulation.

The 2014 October 10 outburst exceeded all preceding records. 
The XRT light curve reached 668 counts s$^{-1}$ when binned at a 
signal-to-noise ratio $S/N=5$, which converts to a peak flux of 
$1.0\times10^{-7}$ erg cm$^{-2}$ s$^{-1}$ (0.3--10\,keV, unabsorbed, 2.1\,Crab),  
or $L\sim 3\times10^{38}$\,erg\,s$^{-1}$, 
when considering a count rate to flux conversion factor  
($1.5\times10^{-10}$ erg cm$^{-2}$ count$^{-1}$) 
derived from the strictly simultaneous 
XRT$+$BAT spectra during interval B (Table~\ref{sfxt14:tab:broadfits}).  

Figure~\ref{sfxt14:fig:lcv_lum} shows the bolometric X-ray 
luminosity light curve of the 2014-10-10 outburst (red)  
compared to the brightest ones recorded by \sw\ for this source. 
The right-hand side y-axis is the standard Eddington luminosity for spherical accretion of fully ionized hydrogen
for a 1.4\,M$_\odot$\ NS according to $L_{\rm Edd}  \simeq1.26 \times 10^{38}  (M/M_\odot)$ erg s$^{-1}$ 
\citep[][we neglected any effect on the Eddington luminosity due to the high NS magnetic 
field as this is not relevant for our conclusions, see Sect.~\ref{sfxt14:discussion}]{BaskoSunyaev75}. 
We also mark the out-of-outburst average count rate obtained 
during the two-year monitoring with \sw/XRT \citep[][]{Romano2011:sfxts_paperVI}, 
$\sim 0.1$  counts s$^{-1}$, 
implying an average flux of 
$\sim1.5\times10^{-11}$ erg cm$^{-2}$ s$^{-1}$ (0.3--10\,keV, unabsorbed)
and bolometric luminosity of $\sim4.7\times10^{34}$ erg s$^{-1}$.

\subsection{Pulsations} \label{sfxt14:pulsations} 

For the timing analysis, we concentrated on the WT mode (time resolution: 1.78\,ms) 
XRT data of the first orbit, since this segment is an uninterrupted stretch of data and 
contains most of the counts (more than 50\,\% of the total). 
In the Fourier transform of the data, a peak ($\nu/\Delta\nu\sim3$) 
arises around 0.086~Hz ($\sim$12~s, see Fig.~\ref{sfxt14:fig:powspec}, left). 
The period of this candidate modulation, as measured by a $Z^2$ test, is $11.58\pm0.03$~s. 
We estimated for this timing feature a quality factor \citep[][]{Klis2000}, 
of $Q=\nu/\Delta\nu\sim3$. 
Although this would be comparable to the typical quality factors of quasi-periodic oscillations 
\citep[QPOs, see, e.g.,][and references therein]{Klis2006}, 
the low Fourier resolution for frequencies around 0.1\,Hz (see Fig.\,\ref{sfxt14:fig:powspec}, left) 
would not allow us to distinguish the case of a QPO from that of a coherent pulsation. 
Given the apparent modulation of the source lightcurve at a period similar to the peak 
of the timing feature, we tentatively ascribe the latter to the pulsation of the NS hosted in \igr.\
The corresponding pulse profile is sinusoidal (Fig.\,\ref{sfxt14:fig:powspec}, right), 
with an RMS pulsed fraction of $9.1\pm0.6\%$ (or $10\pm1\%$ if the pulsed fraction is 
derived using a sinusoidal fit). 
This feature  cannot be tracked down in the following WT or PC data and is embedded 
in a noisy spectrum, making it difficult to evaluate its significance. 

\setcounter{figure}{3}  
\begin{figure}
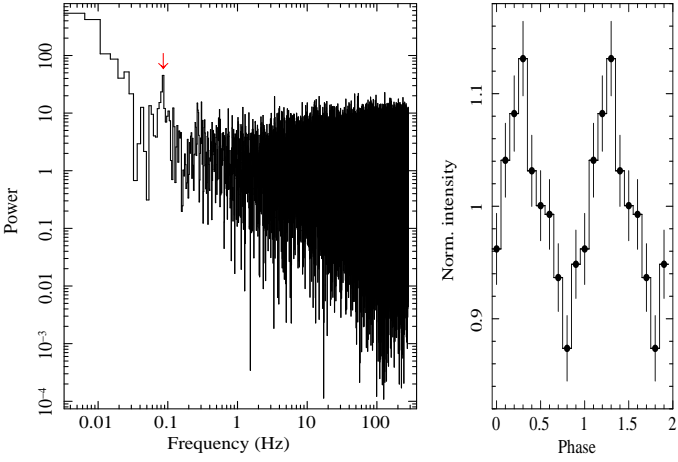

\hspace{-0.3truecm}
  \centerline{\includegraphics[width=6cm,height=6.0cm,angle=270]{figure4a.ps}
\hspace{-0.2truecm}
\includegraphics[width=6cm,height=3.5cm,angle=270]{figure4b.ps}
}
 \caption[XRT light curve]{Left: Fourier power spectrum of the Swift/XRT WT mode data (0.3--10~keV) 
collected during the beginning of the outburst. The peak
corresponding to the 0.086~Hz signal is indicated by the red arrow. 
Right: Folded profile obtained by folding the WT at 
the 11.6-s candidate periodicity (see Sect.~\ref{sfxt14:pulsations} for details).
 		\label{sfxt14:fig:powspec} 
}
\end{figure}

We carefully inspected the light curve to check if the $\sim$12-s signal is due 
to a modulation present along the whole dataset or, given the relatively short exposure, 
it is due to the occurrence of one or few burst-like events.
As shown in Fig.~\ref{sfxt14:fig:pulsations}, the light curve is characterised by strong aperiodic 
variability (which is the source of the red noise in the Fourier power spectrum) and by a 
number of superimposed peaks and dips, which are almost equidistant from each other, 
separated by about 11--12\,s (the origin of the signal). 
These features are present during the greatest part of, but not all, the first orbit data. 
To evaluate the significance of the $\sim$12-s modulation, we fit the light curve with a
model consisting of a constant plus two Gaussians (centred at about 8 and 40~s; the one at
$\sim$8~s has a negative normalisation to account for the large dip around 5\,s) for
the continuum, and added a sinusoidal component to account for the modulation.
All parameters were left free to vary, and we obtained a value of  $P=11.60\pm0.13$\,s (1$\sigma$) 
for the modulation period. 
We applied the F-test to assess the need for including the additional sinusoidal component.
We found that the probability that the inclusion is significant is of 99.984\%, corresponding 
to a detection at about 4$\sigma$ confidence level single-trial (period). 

We also searched for pulsations in the data of the previous bright outbursts that 
afforded long observations and good counting
statistics         
(2011-03-24,  
2013-06-28,   
2013-09-11,   
2014-05-25)   
but no significant signal was
found either around 0.086~Hz or at different frequencies. 
In all cases we obtained 3\,$\sigma$ upper limits on the source pulse fraction of 
$\sim$10\% for a coherent signal in the range of 0.005--20~s.

\setcounter{figure}{4}  
\begin{figure}
   \centering
\hspace{-1.truecm}
 \includegraphics[width=5.5cm,height=9cm,angle=270]{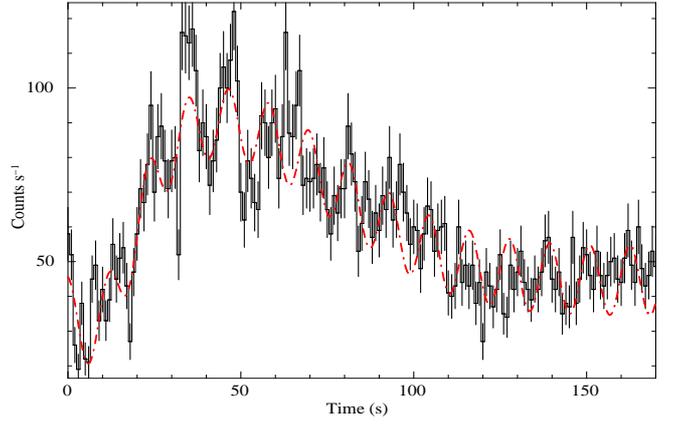}
 \caption[XRT light curve]{\sw/XRT WT (0.3--10~keV) light curve (black)  fitted with a 
model consisting of a constant plus two Gaussian for 
the continuum, and an additional sinusoidal component to account for the modulation.
 		\label{sfxt14:fig:pulsations} }
\end{figure}

\section{Discussion and conclusions\label{sfxt14:discussion}}

In the past few years SFXTs, and \igr\ in particular with its uniquely high dynamic 
range now reaching $\gtrsim$10$^6$, have been challenging our understanding 
of wind accretion onto NSs. 
In all models proposed so far to interpret the peculiar behaviour of these sources 
in X-rays, a particular effort was dedicated to finding  a mechanism to inhibit accretion 
and provide a reliable explanation for their unusually low averaged X-ray luminosity 
compared to classical systems. While different inhibition mechanisms 
could be considered to explain luminosities as low as $\sim$10$^{32}$~erg~s$^{-1}$, 
in all cases it has been assumed that the maximum achievable luminosity is that 
corresponding to the so-called ``direct-accretion regime'', where  
all inhibiting mechanisms are overcome and the Bondi-Hoyle accretion rate sets in  
\citep[see e.g.,][for a recent review]{Shakura2012:quasi_spherical}. 
As we show below, the peak X-ray luminosity of  \igr\ during the exceptional outburst 
reported in this paper reached the standard Eddington limit expected for a NS of 
1.4~$M_{\odot}$ and challenged for the first time the maximum (rather than the minimum) 
theoretical luminosity expected for an SFXT. 

In the classical model of wind accretion, the maximum NS accretion rate is regulated 
by the accretion radius, $R_{\rm acc}\simeq 2GM_{\rm NS}/v_{\rm rel}^2$, 
where $M_{\rm NS}$ is the NS mass and $v_{\rm rel}^2$=$v_{\rm orb}(a)^2$+$v_{\rm w}(a,t)^2$. 
In the latter expression, $v_{\rm orb}(a)$ is the NS orbital velocity at a separation $a$ from 
its companion and $v_{\rm w}(a,t)$ is the companion wind velocity at the NS location and time $t$. 
The rate at which the wind material is captured and accreted by the NS is thus given by 
$\dot{M}_{\rm acc}$=$\pi \, R_{\rm acc}^2 \, v_{\rm rel} \, \rho_{\rm w}(a,t)$ (where $\rho$ is the 
wind density), and the correspondingly released X-ray luminosity can be estimated as 
$L_{\rm X}$=$GM\dot{M}_{\rm acc}$/$R_{\rm NS}$.  
Assuming the simplest case of a spherically symmetric wind and neglecting the NS orbital 
velocity, the wind density can be expressed at any time by 
$\rho_{\rm w}(a)$=$\dot{M}_{\rm w}$/(4$\pi \, a^2 \, v_{\rm w}(a)$), where $\dot{M}_{\rm w}$ 
is the wind mass loss rate.  The Bondi-Hoyle mass accretion rate can be expressed as, 
\begin{equation}
L_{\rm X} \approx 5.8\times10^{35}\dot{M}_{-6} \, a_{\rm 5d}^{-2} \,  v_{8}^{-4} \, {\rm erg~s^{-1},} 
\label{eq:lx} 
\end{equation}
where 
$\dot{M}_{-6}=\dot{M}_{\rm w}/(10^{-6} M_{\odot} \, {\rm yr^{-1}})$, $v_{8}=v_{\rm w}/(10^{8}~{\rm cm\,s^{-1}})$. 
In Eq.~\ref{eq:lx} we assumed a circular orbit to express the separation $a$ as a 
function of the total mass ($M_{\rm tot}$) and the orbital period 
($P_{\rm orb}$) of the system: $a_{\rm 5d}$=2.7$\times$10$^{12} \, P_{\rm 5d}^{2/3} \, M_{30}^{1/3}$, 
where $P_{\rm 5d}$ is the orbital period in units of 5\,d and $M_{30}$ is the total mass of the 
system in units of 30~$M_{\odot}$, chosen to qualitatively match the characteristic values of 
\igr\ (see Sect.~\ref{sfxt14:intro}). 

Equation~(\ref{eq:lx}) shows that the peak luminosity achieved by \igr\ during the 
2014-10-10 outburst can only be explained by assuming a very low wind velocity
 ($v_{w}\lesssim 2\times10^{6}\,{\rm cm\,s^{-1}}$) or an 
unlikely high mass loss rate from the star\footnote{The typical velocity of the 
wind from an O-B supergiant can be as high as 1--$2\times10^8$\,cm\,s$^{-1}$, 
and the mass loss rate is usually comprised to within the range 
10$^{-7}$--10$^{-5}$~$M_{\odot}$~yr$^{-1}$ \citep[see, e.g.,][for a recent review]{Puls2008}.} 
($\dot{M}_{w} \gg$10$^{-6} M_{\odot}$\,yr$^{-1}$). 
Considering the first hypothesis, it is known that the photoionization of the wind 
material by the X-rays emitted from the NS can lead to a substantial reduction of the 
stellar wind velocity, especially in bright HMXBs endowed with short orbital periods 
and eccentric orbits \citep[see, e.g.,][and references therein]{Watanabe2006,Ducci2010}. 
 orbital period of \igr\ could be sufficiently short and the orbit eccentric enough 
\citep{Nikolaeva2013} to allow a substantial ionization of the wind material and 
lead to wind velocities as low as the values indicated above (at least during the 
brightest outbursts when the X-ray emission is more intense). 
If this is the case, it is difficult to avoid the formation of a temporary accretion 
disc around the NS in \igr,\ as a disc is expected to be already in place for wind 
velocities $\lesssim$5$\times$10$^7$~cm~s$^{-1}$ 
\citep[see discussion in][]{Ducci2010}. \citet{Fryxell1988} suggested that instabilities 
of inhomogeneous slow winds in HMXBs could lead not only to the rapid formation
of accretion discs, but also to their quick dissipation 
\citep[e.g., EXO\,2030+375, ][]{Taam1988a}. 
On one hand, accretion from a disc could already provide larger mass accretion 
rates than those typical of wind-fed binaries. 
On the other hand, their dissipation was shown to produce sudden increases 
in the mass accretion rates and the appearance of bright X-ray flares (by a factor 
of $\sim$10). 
The typical duration of these flares is $\sim6\,GM_{\rm NS}/v_{\rm rel}^3$, thus compatible 
with the observational properties of the super-outburst from \igr\ (assuming the 
values mentioned above for $v_{\rm rel}$). 
Additional simulations presented by \citet[][]{Taam1988b} also showed that a 
more erratic or periodic behavior of the flares can be obtained by assuming different 
properties of (poorly known) density and velocity asymmetry of the accretion flow.   
The presence of short-lived accretion discs in SFXTs (and other NS HMXBs in general) 
could be tested observationally through the detection of rapid spin-up phases 
\citep{Klochkov2011,Jenke2012}. 
However, \igr\ did not display convincing evidence of pulsations in the past 
(see Sect.~\ref{sfxt14:intro}) and we could only reveal a marginally significant 
pulsed signal at $\sim11.6$~s during the peak of the event reported in this paper. 
No meaningful searches for spin period derivatives can thus be carried out. 

We cannot currently rule out alternative explanations for the exceptionally bright 
outburst of \igr\ recorded by \sw. A viable mechanism to overcome 
the difficulties in the standard Bondi-Hoyle accretion scenario is that of the 
``ingestion of a massive clump'' \citep{Bozzo2011:18410}. 
In this case, it is assumed that the bright outburst was due to a dense structure 
in the wind of the supergiant star being accreted onto the compact object. 
The equation relating the density $\rho_{\rm w}(a,t)$ and the mass loss rate from 
the supergiant introduced above is no longer valid, as massive structures are created 
by definition in asymmetric winds. A sufficiently high increase in the density of 
the accreting material (together with a possible decrease in the corresponding 
velocity) could thus be invoked to achieve sufficiently high $\dot{M}_{\rm acc}$ 
without implying an enhanced $\dot{M}_{\rm w}$. 
Even though current multi-dimensional simulations of winds from massive stars 
do not favour the formation of large and very dense clumps \citep[see the 
discussion in][and references therein]{Bozzo2015:underluminous}, at least in the case 
of IGR~J18410$-$0535 X-ray observations revealed some evidence for the ingestion 
of a dense structure onto the NS \citep{Bozzo2011:18410}. 
The spectroscopic analysis of the outburst from \igr\ did not reveal any change in 
the source spectral properties that could favour such a hypothesis.
 While it could still be assumed that this is due to projection effects between the 
velocity vector of the clump and the observed line of sight, we consider in the present 
case this explanation unlikely. 

The detection of other bright outbursts from the SFXTs with \sw\ and, in the 
future, with the next generation of X-ray facilities providing fine X-ray spectroscopy 
on still unexplored short time scales 
\citep[e.g, the LOFT/LAD;][]{Feroci2014,Orlandini2015},  will help clarify the 
origin of these events.

\begin{acknowledgements}
We wholeheartedly thank the {\it Swift} team duty scientists and science planners 
for their courteous efficiency, and S.D.\ Barthelmy, D.N.\ Burrows, and N.\ Gehrels.
We thank S.E.\ Motta, L.\ Stella, A.\ Beardmore, M.\ Capalbi, and S.\ Vercellone 
for useful discussions. 
We also thank our referee for swift comments that helped improve the paper. 
PR and SC acknowledge contract ASI-INAF I/004/11/0.  
PE acknowledges a Fulbright Research Scholar grant administered by the
U.S.--Italy Fulbright Commission and is grateful to the
Harvard--Smithsonian Center for Astrophysics for hosting him during
his Fulbright exchange. 
LD thanks Deutsches Zentrum f\"ur Luft und Raumfahrt (Grant FKZ 50 OG 1301). 
\end{acknowledgements} 

\bibliographystyle{aa} 

%

\Online


\setcounter{table}{0} 
\begin{table}  
\tiny
 \tabcolsep 2pt         
 \begin{center}         
 \caption{Observation log. 
          \label{sfxt14:tab:alldata} } 
 \begin{tabular}{llllr}
 \hline 
 \hline 
 \noalign{\smallskip} 
Sequence   & Obs/Mode  & Start time& End time   (UT) & Expo.  \\ 
                      &     &   (UT)   &   (UT)   &  (s) \\ 
 \noalign{\smallskip} 
 \hline 
 \noalign{\smallskip} 
00614903000	&	BAT/evt	&	2014-10-10 15:00:30	&	2014-10-10 20:00:06	&	1648 	\\
00614903000	&	XRT/WT	&	2014-10-10 15:06:45	&	2014-10-10 19:15:26	&	420		\\
00614903000	&	XRT/PC	&	2014-10-10 16:08:27	&	2014-10-10 19:56:22	&	7173	\\
00035056161	&	XRT/PC	&	2014-10-11 16:02:14	&	2014-10-11 16:18:45	&	973		\\
00035056162	&	XRT/PC	&	2014-10-12 20:49:24	&	2014-10-12 21:05:51	&	978		\\
00035056163	&	XRT/PC	&	2014-10-13 16:29:14	&	2014-10-13 17:40:37	&	960		\\
00035056165	&	XRT/PC	&	2014-10-14 22:26:08	&	2014-10-14 22:41:54	&	925		\\
00035056166	&	XRT/PC	&	2014-10-15 16:07:04	&	2014-10-15 16:24:55	&	1071	\\
  \noalign{\smallskip}
00449907000$^{\mathrm{a}}$  &   XRT/WT &  2011-03-24 01:59:15    &       2011-03-24 03:18:22     &  657     \\  
00449907000$^{\mathrm{a}}$  &   XRT/PC &   2011-03-24 02:10:07    &       2011-03-24 04:01:03     &  3360   \\
00035056150$^{\mathrm{a}}$  &   XRT/PC &   2011-03-27 19:54:23    &       2011-03-27 20:11:56    &1046    \\
00035056151$^{\mathrm{a}}$  &   XRT/PC &   2011-03-28 20:10:36    &       2011-03-28 20:21:58    &662     \\
00035056152$^{\mathrm{a}}$  &   XRT/PC &   2011-03-29 20:15:13    &       2011-03-29 20:23:57    &504     \\
  \noalign{\smallskip}
00559221000     &       XRT/PC  &       2013-06-28 07:28:56     &       2013-06-28 08:48:24     &       1010    \\
00570402000     &       XRT/WT  &       2013-09-11 16:01:52     &       2013-09-11 17:40:54     &       1024    \\
00570402000     &       XRT/PC  &       2013-09-11 16:18:49     &       2013-09-11 17:59:42     &       1394    \\
  \noalign{\smallskip}
00599954000	&	XRT/WT	&	2014-05-25 22:33:22	&	2014-05-25 22:34:31	&	47	\\
00599955000	&	XRT/WT	&	2014-05-25 22:34:54	&	2014-05-25 22:55:55	&	1235	\\
00035056156	&	XRT/PC	&	2014-05-26 19:03:35	&	2014-05-26 19:20:05	&	978	\\
00035056157	&	XRT/PC	&	2014-05-27 16:15:36	&	2014-05-27 16:30:56	&	920	\\
00035056158	&	XRT/PC	&	2014-05-30 16:16:27	&	2014-05-30 22:38:53	&	1241	\\
00035056159	&	XRT/PC	&	2014-05-31 19:11:56	&	2014-05-31 21:00:54	&	1168	\\
00035056160	&	XRT/PC	&	2014-06-01 09:27:33	&	2014-06-01 21:09:54	&	1133	\\
  \noalign{\smallskip}
  \hline
  \end{tabular}
  \end{center}
  \begin{list}{}{} 
  \item[$^{\mathrm{a}}$] See \citep[][]{Farinelli2012:sfxts_paperVIII}. 
  \end{list} 
  \end{table} 

\setcounter{table}{1} 
\begin{table} 	
  \tabcolsep 4pt         
 \begin{center} 	
 \caption{Spectral fits of simultaneous XRT/WT and BAT data.} 	
 \label{sfxt14:tab:broadfits} 	
 \begin{tabular}{ cccccc ccc} 
 \hline 
 \hline 
 \noalign{\smallskip} 
 Spectrum\tablefootmark{a}	&Time  &	$N_{\rm H}$ &$\Gamma$ & $E_{\rm fold}$ & $E_{\rm cut}$    &Flux\tablefootmark{b} & Luminosity\tablefootmark{c} & $\chi^{2}_{\nu}$/dof  \\
              & (s since $T_0$)  & ($\times10^{22}$~cm$^{-2}$) &   &(keV)  & (keV)  &  ($\times10^{-8}$ erg cm$^{-2}$ s$^{-1}$)  &  ($\times10^{37}$ erg s$^{-1}$)   &    \\
           & 	         &                                       & 	                   &  	          &    	              &  (0.3--10\,keV)           &  (0.01--1000\,keV)   &    \\
 \noalign{\smallskip} 
 \hline 
 \noalign{\smallskip} 
Orb 1  &136.2--306.8     &$1.31_{-0.10}^{+0.11}$ &    $0.72_{-0.07}^{+0.07}$ &    $11.9_{-1.0}^{+1.0}$ &     $6.7_{-0.4}^{+0.4}$ &$3.1_{-0.1}^{+0.1}$  &$10.0$ & $1.01/364$	\\
 \noalign{\smallskip} 
A	&136.2--166.2	&$1.37_{-0.29}^{+0.34}$ &    $0.70_{-0.20}^{+0.21}$ &    $11.4_{-1.9}^{+2.0}$ &     $6.6_{-0.7}^{+0.8}$ &$2.7_{-0.1}^{+0.1}$ &$8.8$  & $0.92/83$ 	\\
B      &166.2--186.2        &$1.51_{-0.25}^{+0.28}$ &    $0.85_{-0.17}^{+0.18}$ &    $11.2_{-1.9}^{+1.9}$ &     $6.7_{-0.7}^{+0.9}$ &$4.8_{-0.2}^{+0.2}$ &$15.3$& $0.84/99$	\\
Peak\tablefootmark{d} &169.6--172.8       & --& -- & -- & -- &$10.0$ &$32.0$&  --	\\
C	&186.2--206.2	&$1.35_{-0.26}^{+0.29}$ &    $0.71_{-0.19}^{+0.20}$ &    $11.1_{-1.9}^{+1.9}$ &     $6.7_{-0.7}^{+0.8}$ &$4.1_{-0.2}^{+0.2}$ &$13.2$ & $1.10/89$	\\
D	&206.2--226.2	&$1.51_{-0.30}^{+0.35}$ &    $0.86_{-0.21}^{+0.22}$ &    $11.7_{-2.4}^{+2.5}$ &     $7.6_{-1.0}^{+1.2}$ &$3.3_{-0.2}^{+0.2}$ &$10.7$ & $1.06/77$ 	\\
E	&226.2--246.2	&$1.32_{-0.26}^{+0.31}$ &    $0.77_{-0.19}^{+0.21}$ &    $ 9.1_{-2.5}^{+2.4}$  &     $7.9_{-0.9}^{+1.1}$ &$3.3_{-0.2}^{+0.2}$ &$10.6$ & $0.98/72$       \\
F	&246.2--266.2	&$0.88_{-0.29}^{+0.35}$ &    $0.48_{-0.22}^{+0.24}$ &    $10.3_{-2.3}^{+2.3}$ &     $6.7_{-0.8}^{+1.0}$ &$2.3_{-0.1}^{+0.1}$ &$7.3$ & $0.98/56$ 	\\
G	&266.2--286.2	&$1.03_{-0.31}^{+0.37}$ &    $0.51_{-0.23}^{+0.25}$ &    $10.5_{-2.4}^{+2.2}$ &     $6.2_{-0.8}^{+1.0}$ &$2.5_{-0.2}^{+0.2}$  &$8.0$& $0.89/58$  \\
H	&286.2--306.8	&$0.83_{-0.24}^{+0.30}$ &    $0.55_{-0.20}^{+0.22}$ &    $ 9.7_{-2.2}^{+2.1}$  &     $6.0_{-0.9}^{+1.1}$ & $2.5_{-0.1}^{+0.1}$  &$8.0$ & $0.99/62$  	\\
  \noalign{\smallskip}
  \hline
  \end{tabular}
  \end{center}
\tablefoot{
\tablefoottext{a}{Defined in Sect.~\ref{sfxt14:superburst}, see Fig.~\ref{sfxt14:fig:lcvs}c.}
\tablefoottext{b}{Average unabsorbed 0.3--10\,keV fluxes in units of 10$^{-8}$~erg~cm$^{-2}$~s$^{-1}$.}
\tablefoottext{c}{Bolometric luminosity in units of 10$^{37}$~erg~s$^{-1}$.}
\tablefoottext{d}{The peak luminosity, calculated by converting peak count rate of 668 counts s$^{-1}$ by using the spectral parameters of spectrum B.}
}
  \end{table} 

\setcounter{figure}{1}  

\begin{figure}
\hspace{-0.3truecm}
  \centerline{\includegraphics[width=6cm,height=9cm,angle=270]{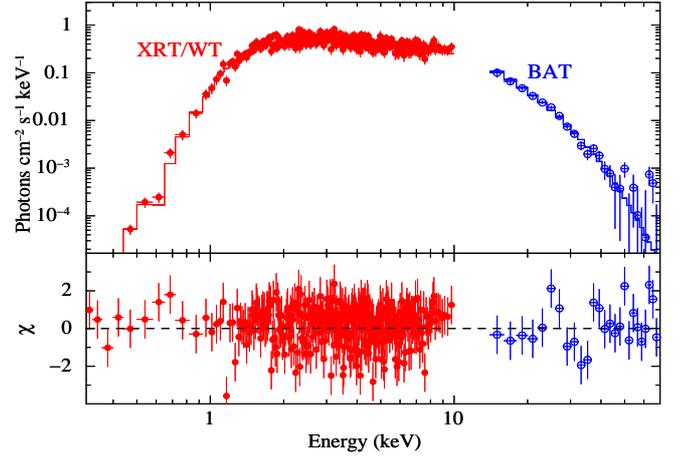}}
 \caption[XRT light curve]{Spectroscopy of the 2014-10-10 outburst of \igr. 
Top panel: simultaneous XRT/WT (filled red circles) and BAT (empty blue circles) data fit with 
an absorbed  power law with a high energy exponential cut-off  model. 
Bottom panel: the residuals of the fit (in units of standard deviations).
 		\label{sfxt14:fig:broadspec} 
}
\end{figure}

\end{document}